\documentclass[nofootinbib,prl,reprint,longbibliography]{revtex4-1}

\usepackage{hyperref}
\usepackage{graphicx}
\usepackage{verbatim}
\usepackage{times}
\usepackage{bm}
\usepackage{color}
\usepackage{amsmath}
\usepackage{hyperref}

\newcommand\jcap{{J. Cosmology Astropart. Phys.}}
\newcommand\na{{New A.}}%

\makeatletter
\def\ignorespacesandimplicitepars{%
  \begingroup
  \catcode13=10
  \@ifnextchar\relax
    {\endgroup}%
    {\endgroup}%
}

\let\jnl@style=\rm
\def\ref@jnl#1{{\jnl@style#1}}

\def\aj{\ref@jnl{AJ}}                   
\def\araa{\ref@jnl{ARA\&A}}             
\def\apj{\ref@jnl{ApJ}}                 
\def\apjl{\ref@jnl{ApJ}}                
\def\apjs{\ref@jnl{ApJS}}               
\def\ao{\ref@jnl{Appl.~Opt.}}           
\def\apss{\ref@jnl{Ap\&SS}}             
\def\aap{\ref@jnl{A\&A}}                
\def\aapr{\ref@jnl{A\&A~Rev.}}          
\def\aaps{\ref@jnl{A\&AS}}              
\def\azh{\ref@jnl{AZh}}                 
\def\baas{\ref@jnl{BAAS}}               
\def\jrasc{\ref@jnl{JRASC}}             
\def\memras{\ref@jnl{MmRAS}}            
\def\mnras{\ref@jnl{MNRAS}}             
\def\pra{\ref@jnl{Phys.~Rev.~A}}        
\def\prb{\ref@jnl{Phys.~Rev.~B}}        
\def\prc{\ref@jnl{Phys.~Rev.~C}}        
\def\prd{\ref@jnl{Phys.~Rev.~D}}        
\def\pre{\ref@jnl{Phys.~Rev.~E}}        
\def\prl{\ref@jnl{Phys.~Rev.~Lett.}}    
\def\pasp{\ref@jnl{PASP}}               
\def\pasj{\ref@jnl{PASJ}}               
\def\qjras{\ref@jnl{QJRAS}}             
\def\skytel{\ref@jnl{S\&T}}             
\def\solphys{\ref@jnl{Sol.~Phys.}}      
\def\sovast{\ref@jnl{Soviet~Ast.}}      
\def\ssr{\ref@jnl{Space~Sci.~Rev.}}     
\def\zap{\ref@jnl{ZAp}}                 
\def\nat{\ref@jnl{Nature}}              
\def\iaucirc{\ref@jnl{IAU~Circ.}}       
\def\aplett{\ref@jnl{Astrophys.~Lett.}} 
\def\apspr{\ref@jnl{Astrophys.~Space~Phys.~Res.}}
\def\bain{\ref@jnl{Bull.~Astron.~Inst.~Netherlands}} 
\def\fcp{\ref@jnl{Fund.~Cosmic~Phys.}}  
\def\gca{\ref@jnl{Geochim.~Cosmochim.~Acta}}   
\def\grl{\ref@jnl{Geophys.~Res.~Lett.}} 
\def\jcp{\ref@jnl{J.~Chem.~Phys.}}      
\def\jgr{\ref@jnl{J.~Geophys.~Res.}}    
\def\jqsrt{\ref@jnl{J.~Quant.~Spec.~Radiat.~Transf.}}
\def\memsai{\ref@jnl{Mem.~Soc.~Astron.~Italiana}}
\def\nphysa{\ref@jnl{Nucl.~Phys.~A}}   
\def\physrep{\ref@jnl{Phys.~Rep.}}   
\def\physscr{\ref@jnl{Phys.~Scr}}   
\def\planss{\ref@jnl{Planet.~Space~Sci.}}   
\def\procspie{\ref@jnl{Proc.~SPIE}}   

\makeatother

\newcommand{\prlsection}[1]{\emph{#1.}---\ignorespacesandimplicitepars}

\newcommand{\blue}{}

\newcommand{\prv}[1]{\bm{\mathrm{#1}}}
\newcommand{\pru}[1]{\bm{\hat\mathrm{#1}}}
\newcommand{\vecx}{\prv{x}}
\newcommand{\veck}{\prv{k}}
\newcommand{\vecK}{\prv{K}}
\newcommand{\vecl}{\prv{\ell}}
\newcommand{\p}{\phantom{a}}

\newcommand{\hp}{\mathcal{H}^\perp}
\newcommand{\Pp}{\mathcal{P}}
\newcommand{\transfer}{\mathcal{T}}
\DeclareMathOperator{\Cov}{Cov}
\newcommand{\heliP}{\Delta\chi}

\begin{document}

\title{Two- and Three-Dimensional Probes of Parity in Primordial Gravity Waves}

\author{Kiyoshi Wesley Masui}
    \email[]{kiyo@physics.ubc.ca}
    \affiliation{Department of Physics and Astronomy, University of British 
        Columbia, 6224 Agricultural Rd, Vancouver, British Columbia, V6T 1Z1, Canada}
\author{Ue-Li Pen}
    \affiliation{Canadian Institute for Theoretical Astrophysics, 60 St George Street, Toronto, Ontario M5S 3H8, Canada}
    \affiliation{Canadian Institute for Advanced Research, CIFAR Program in Cosmology and Gravity, Toronto, ON, M5G 1Z8}
    \affiliation{Dunlap Institute for Astronomy \& Astrophysics, University of Toronto, 50 St George St, Toronto, ON, M5S 3H4, Canada}
    \affiliation{Perimeter Institute for Theoretical Physics, Waterloo, 
    Ontario N2L 2Y5, Canada}
\author{Neil Turok}
    \affiliation{Perimeter Institute for Theoretical Physics, Waterloo, 
    Ontario N2L 2Y5, Canada}

\date{\today}

\begin{abstract}

We show that three-dimensional information is critical to discerning the effects
of parity violation in the primordial
gravity-wave background. If present, helical gravity waves induce parity-violating correlations in the cosmic microwave background
(CMB) between
parity-odd
polarization $B$-modes and parity-even temperature anisotropies ($T$) or polarization
$E$-modes.
Unfortunately, $EB$ correlations are much
weaker than would be naively expected, which we show
is due to an approximate symmetry resulting from the
two-dimensional nature of the CMB. The detectability of parity-violating
correlations is
exacerbated by the fact that the handedness of individual modes cannot be
discerned in the two-dimensional CMB, leading to a noise contribution
from scalar matter
perturbations.
In contrast, the tidal imprints of primordial gravity waves fossilized
into the large-scale structure of the Universe are a three-dimensional
probe of parity violation. Using such fossils the handedness
of gravity waves may be determined on a mode-by-mode basis,
permitting future surveys to probe helicity at the percent level if the
amplitude of primordial gravity waves is near current observational upper
limits.

\end{abstract}

\maketitle


Nature is parity violating: the electroweak $W$ bosons couple only
to left-handed particles and right-handed antiparticles, violating both parity
($P$) and charge conjugation ($C$) maximally.
Thus, weak nuclear processes produce only left-handed
neutrinos and right-handed antineutrinos. In this context, it is natural to ask
whether gravitational processes violate parity, in particular, whether such
violations may be present in the cosmological gravity-wave background.
If detected, the stochastic background of long wavelength gravity waves would
provide a uniquely powerful probe of the very early Universe.
A variety of sources of gravitational parity violation have been
considered, from fundamental quantum gravity
effects to
rolling inflationary
axions~\citep{1999PhRvL..83.1506L,2008PhRvL.101n1101C,
2009PhRvL.102w1301T,2011JCAP...06..003S}.
Each of these could
have have left an imprint on the net helicity of the gravity wave background,
namely the preferred excitation of one circular polarization over the
other.

An additional reason to be interested in a helical primordial
gravity-wave background is that it would have brought
nonperturbative standard model processes into play, potentially explaining the
cosmological matter--antimatter
asymmetry~\cite{Alexander:2004us}. Because of the chiral nature of
neutrinos, lepton number conservation is violated in the standard model by a
gravitational anomaly: $\partial_\mu J^\mu_L=(3/32 \pi^2) \varepsilon^{\alpha
\beta \gamma \sigma} R_{\alpha \beta}^{\quad \rho \sigma} R_{\rho \sigma \gamma
\sigma}$, where $J^\mu_L$ is the lepton current and $R_{\alpha\beta\rho\sigma}$
is the Riemann curvature tensor~\cite{AlvarezGaume:1983ig}.
Any process that generates a
parity-violating ensemble of gravity waves will necessarily give the
anomaly a nonzero expectation value, driving a lepton asymmetry number that
would, at high temperature, be converted into a baryon asymmetry by electroweak
``sphaleron'' processes.

The net helicity of the gravity-wave background is thus of fundamental
interest and importance. In this Letter, we consider its
detectability, and the role played by the dimensionality of potential probes.
Gravitational waves induce both CMB temperature and polarization anisotropies
\cite{1997PhRvL..78.2054S, 1997PhRvL..78.2058K}. While temperature and $E$-mode
polarization patterns are also induced by scalar perturbations, at
linear order $B$-mode polarization is only produced by gravity waves.
Cross-correlations between the parity-odd $B$-mode polarization and
either the parity-even $E$-mode polarization or temperature perturbations are
parity violating. These correlations vanish if the gravity wave background has
no net helicity, as is the case in standard inflationary scenerios.

\citet{2007JCAP...09..002S} calculated the resulting $EB$ and $TB$ correlations for
a primordial gravity-wave background with net helicity. They found that even
in the maximally helical case---with all gravity waves having the same circular
polarization---$EB$ correlations are highly suppressed, with cross-correlation
coefficients of order $\sim 10^{-2}$. The $TB$ correlations are larger, with
correlation coefficients closer to unity, but these are difficult to detect
against the large background of noise arising from the scalar matter
perturbations.
Detecting helicity in the CMB is thus very difficult:
even for a cosmic-variance limited
measurement of the CMB polarization, parity violation is never detectable at
$>3\sigma$ if the amplitude of primordial gravity waves---parameterized by the
tensor-to-scalar ratio $r$---is less than $0.02$ \citep{2010PhRvD..81l3529G}.
Even at the current upper limit of $r < 0.07$ \citep{2016PhRvL.116c1302B},
only fractional helicity power above $\Delta\chi = 50\%$ could be detected. Indeed a
recent search for such correlations concluded that there is little hope for
their detection in current or upcoming CMB surveys \citep{2016JCAP...07..044G}.

After introducing some formalism for gravity wave helicity, we show that the
suppression of the $EB$ correlations is a direct result of the two-dimensional
nature of the CMB.
We then consider whether helicity can be detected using the three-dimensional
tidal imprints of gravity waves fossilized into the large-scale structure,
as introduced by
\citet{2010PhRvL.105p1302M}. We show that the limitations of the CMB are
removed in three dimensions, in principle permitting future surveys to detect
even weak helicity.

\prlsection{Helical gravity waves}

Gravity waves (which we will also refer to as tensor modes)
are encoded in the transverse, traceless part of the linearized
metric perturbations $h_{ab}(\vecx)$, which has two degrees of freedom.
In Fourier space,
$h_{ab}(\vecK) = \int d^3\vecx e^{-i\vecK\cdot\vecx}h_{ab}(\vecx)$, these two
degrees of freedom can be described by the circular polarization amplitudes
$h_R$ and $h_L$ as
\begin{align}
    h_{ab}(\vecK) &= h_R(\vecK)e^R_{ab}(\pru{K}) + h_L(\vecK)
    e^L_{ab}(\pru{K})
    \nonumber\\
    e^R_{ab}(\pru{K}) &\equiv \frac{1}{\sqrt{2}} \left[e^+_{ab}(\pru{K})
+ ie^\times_{ab}(\pru{K})\right]
    \nonumber\\
    e^L_{ab}(\pru{K}) &\equiv \frac{1}{\sqrt{2}} \left[e^+_{ab}(\pru{K})
- ie^\times_{ab}(\pru{K})\right]
    ,
\end{align}
where
$
e^+_{ab}(\pru{K}) \equiv \hat{e}^{\theta_K}_a\hat{e}^{\theta_K}_b
    - \hat{e}^{\phi_K}_a\hat{e}^{\phi_K}_b
$
and
$
    e^\times_{ab}(\pru{K}) \equiv \hat{e}^{\theta_K}_a\hat{e}^{\phi_K}_b
    + \hat{e}^{\phi_K}_a\hat{e}^{\theta_K}_b
$, and $\pru{e}^{\theta_K}$ and $\pru{e}^{\phi_K}$ are the spherical polar unit
vectors relative to $\vecK$.
This definition is consistent with the International Astronomical Union's
definition of circular polarization for electromagnetic radiation, where a
right-hand circular wave's instantaneous spatial configuration forms a
left-handed screw \citep{1996A&AS..117..161H}.

Translational invariance of the statistical ensemble dictates that the only
nonzero correlators have zero net momentum: for this reason only
$h_{ab}(\vecK)$ and $h_{ab}(-\vecK)=h_{ab}(\vecK)^*$ can be correlated in a
statistically homogeneous ensemble. The left- and right-handed circular modes
have opposite helicity, meaning that they transform as $e^{\pm 2 i \alpha}$
under rotations by $\alpha$ about $\vecK$. It follows that the only rotational
invariant (helicity zero) combinations of $h_{ab}(\vecK)$ and $h_{ab}(-\vecK)$
are $LL$ or $RR$ (in obvious notation). These correlators are perfectly
consistent with homogeneity and isotropy, although they individually violate
parity. Any $RL$ correlation is prohibited by the combination of statistical
homogeneity and isotropy.
 
The surviving correlations can be written in terms of power spectra:
\begin{align}
    \langle &h_{ab}(\vecK) h_{cd}(\vecK')\rangle
    =
    (2 \pi)^3 \delta^3(\vecK + \vecK')
    \nonumber\\
    &\phantom{=}
    \times
    \left[
    e^R_{ab}(\pru{K})e^R_{cd}(-\pru{K})
    P_R(K)
    + e^L_{ab}(\pru{K})e^L_{cd}(-\pru{K})
    P_L(K)
    \right],
\end{align}
where $\delta^3(\vecK)$ is the three-dimensional Dirac delta function.
This can be rewritten
\begin{align}
    \langle &h_{ab}(\vecK) h_{cd}(\vecK')\rangle
    =
    (2 \pi)^3 \delta^3(\vecK + \vecK')
    \frac{P_h(K)}{4}
    \nonumber\\
    &\phantom{=}
    \times
    \left\{
    \left[
    e^R_{ab}(\pru{K})e^R_{cd}(-\pru{K})
    + e^L_{ab}(\pru{K})e^L_{cd}(-\pru{K})
    \right]
    \right.
    \nonumber\\
    &\phantom{=}
    \left.
    +
    \heliP(K)
    \left[
    e^R_{ab}(\pru{K})e^R_{cd}(-\pru{K})
    - e^L_{ab}(\pru{K})e^L_{cd}(-\pru{K})
    \right]
    \right\},
\end{align}
where we have defined the tensor power spectrum
$P_h \equiv 2(P_R + P_L)$ and the fractional helicity spectrum
$\heliP \equiv 2(P_R - P_L) / P_h$, with $|\heliP| = 1$ corresponding to
maximal helicity. The factor of $2$ is present by convention
\citep{2009ApJS..180..330K}.

The tensor expressions in square brackets lie in the plane perpendicular to
$\vecK$ and are rotationally invariant within this plane,
as required by
statistical isotropy.  It is
always possible to rewrite a rotationally invariant tensor in terms of the
two-dimensional Kronecker delta, $\delta_{ab} - \hat K_a\hat K_b$ and the
two-dimensional Levi-Civita symbol, $\hat K^c\varepsilon_{cab}$.
Inserting the definitions of the polarization tensors, the above expression
can be reduced to
\begin{widetext}
\begin{align}
    \langle h_{ab}(\vecK) h_{cd}(\vecK')\rangle
    &=
    (2 \pi)^3 \delta^3(\vecK + \vecK') 
    \frac{P_h(K)}{4} 
    \left[
        w_{abcd}(\pru{K})
        +
        \heliP(K) v_{abcd}(\pru{K})
    \right]
    \nonumber\\
    w_{abcd}(\pru{K})
    &\equiv
    (\delta_{ad} - \hat K_a\hat K_d)(\delta_{bc} - \hat K_b\hat K_c)
    + (\delta_{ac} - \hat K_a\hat K_c)(\delta_{bd} - \hat K_b\hat K_d)
    - (\delta_{ab} - \hat K_a\hat K_b)(\delta_{cd} - \hat K_c\hat K_d)
\nonumber\\
v_{abcd}(\pru{K})
    &\equiv
    -i \left[
    (\delta_{ac} - \hat K_a \hat K_c)\hat K_e\varepsilon^e_{\p bd}
    +\hat K_f\varepsilon^f_{\p ac}(\delta_{bd} - \hat K_b \hat K_d)
    \right].
\end{align}

\end{widetext}

The two power spectra $P_h(K)$ and $\heliP(K)$ can be isolated through
contractions of these two point functions \citep{2009A&A...495..697W}:
\begin{align}
    &\langle h^{ab}(\vecK) h_{ab}(\vecK')\rangle 
    = (2\pi)^3\delta^3(\vecK+\vecK')P_h(K)
    \nonumber\\
    &\langle  h^{ab}(\vecK)(-i \hat K^d \varepsilon_{da}^{\p\p c}) h_{cb}(\vecK') \rangle
    \nonumber\\
    &\qquad\qquad= (2\pi)^3\delta^3(\vecK + \vecK')\heliP(K)P_h(K).
    \label{e:contractions}
\end{align}
Here $-i \hat K^c \varepsilon_{ca}^{\p\p b}$ acts as the gravity wave
helicity operator,
noting that \citep{2007JCAP...09..002S}
\begin{align}
    -i \hat K^c \varepsilon_{ca}^{\p\p d} e^R_{db} = e^R_{ab},
    \qquad
    -i \hat K^c \varepsilon_{ca}^{\p\p d} e^L_{db} = -e^L_{ab}.
\end{align}

\begin{figure}
    \includegraphics[scale=0.22]{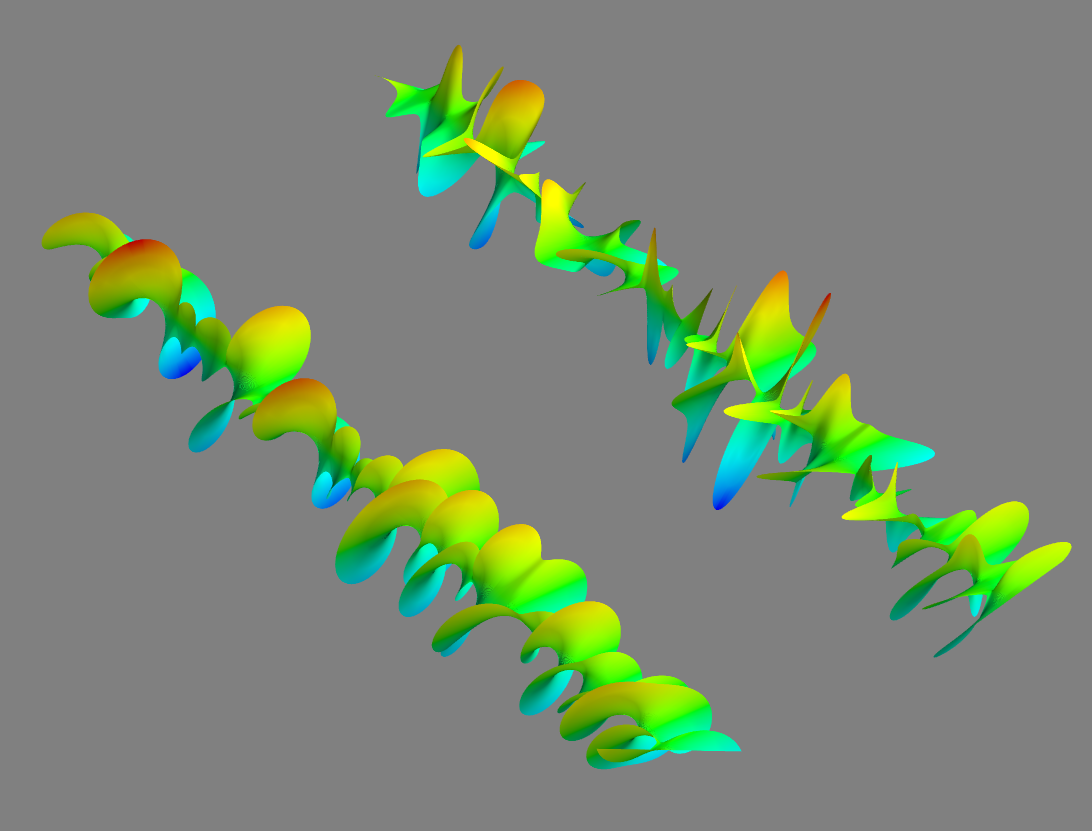}
    \caption{
        \label{f:visualize}
        Visualization of gravity-wave helicity. Shown is the instantaneous
        strain for two pseudo-1D tensor
        fields with strain in the $x-y$ plane and wave numbers aligned with
        the $z$
        axis. The width and orientation of the ribbons correspond to magnitude
        and orientation of the strain. The two fields are generated from the
        same realization of random numbers but the lower field has all power in
        the right-hand circular polarization ($\heliP=1$),
        whereas the power in the upper
        field is
        distributed equally between left- and right-hand circular
        polarizations ($\heliP=0$).
        The realization is drawn from a band-limited white
        power spectrum.
    }
\end{figure}

Figure~\ref{f:visualize} shows a visualization of a Gaussian random gravity-wave
field with and without helicity. The qualitative difference between a helical
and nonhelical field when observed in three dimensions is obvious, while on any
planar slice perpendicular to $\pru K$ the fields are
statistically indistinguishable. For nonperpendicular slicings the fields are
distinguishable since the slicing will cut through phase fronts,
but become indistinguishable once all orientations of $\pru K$
are superimposed. This is in essence why detecting helicity with a two
dimensional probe is difficult, as we now show explicitly for the CMB.

\prlsection{Cosmic microwave background}

If primordial gravity waves had net helicity, one might naively expect the $EB$
angular power spectrum to be $C_\ell^{EB} \sim \heliP
(C_\ell^{BB}C_\ell^{(T),EE})^{1/2}$, where the superscript $(T)$ indicates that
only the contribution from tensor perturbations is included. 
Indeed $C_\ell^{TB} \sim \heliP
(C_\ell^{BB}C_\ell^{(T),TT})^{1/2}$ roughly holds. Instead
\citet{2007JCAP...09..002S} showed that $C_\ell^{EB}$ is two orders of
magnitude smaller than this expectation, making these correlations undetectable.
The $TB$ correlations are hard to detect since the temperature anisotropies get a
large contribution from scalar matter perturbations (especially above
$\ell\sim50$) and the error on the correlations is 
$\Delta C_\ell^{TB} \propto (C_\ell^{TT}C_\ell^{BB})^{1/2}$, \textit{i.e.}
including the contribution from scalar perturbations. That scalar perturbations
induce noise in this correlation is itself a consequence of the loss of
geometric information in the 2D projection.

We now show that the
suppression of $C_\ell^{EB}$ is due to the two-dimensional nature of the CMB.
CMB polarization is generated by the gravity-wave strain projected onto the
two-dimensional surface of last scattering \citep{1997NewA....2..323H}.
This strain generates a local
quadrupole moment in the photon distribution seen by electrons at last scatter,
which perturbs the observed CMB radiation through Thompson scattering. The
strain in the line-of-sight direction has no effect on the polarization,
since the cross-section for
colinear and antilinear scattering vanishes. Like polarization, the projected
strain is a two-dimensional, rank-2 tensor, which can be decomposed into a
convergence, $E$-mode shear and $B$-mode shear. These induce temperature
anisotropies, $E$-mode
polarization, and $B$-mode polarization respectively.

In the absence of parity symmetry, the approximate reflective symmetry
about the surface
of last scattering forces the $EB$ correlations to nearly vanish. For an
approximately flat-sky patch of the CMB, an observer on the
other side of the last scattering surface would see an identical projected
strain (and thus CMB polarization) as we would, except reflected. The
reflection causes a sign flip to only the observed $B$-modes and thus that
observer would measure the opposite $EB$ correlations as we do. Since
that observer lives in the same Universe that we do, they expect the same
power spectra and thus the $EB$ power spectrum must vanish for a flat sky.
\blue{This is true independent of the source of the polarization-generating
photon quadrupole, and applies for helical tensor and vector modes.}

To show this formally, we derive the $EB$ power spectrum within the flat-sky
approximation. The detailed derivation is given in the Supplementary Material
and here we present an abbreviated version. We write the
contribution to the projected strain from a single mode with wave vector $\vecK$
as
\begin{align}
    \hp_{ab}(\pru n,\vecK,\chi)
    &\equiv
    \left[
    \delta^{\perp}_{ac} \delta^{\perp}_{bd}
    -\frac{1}{2} \delta^{\perp}_{ab} \delta^{\perp}_{cd}
    \right]
    e^{i\chi \pru n \cdot\vecK}
    h^{cd}(\vecK)
    .
    \label{e:projected_strain}
\end{align}
Here $\delta^{\perp}_{ab} \equiv \delta_{ab} - \hat n_a \hat n_b$ and
$\chi$ is the comoving radial distance. The
expression in square brackets serves to project the strain into the plane
perpendicular to the line of sight (since the line-of-sight component does not
induce polarization), and removes the trace to isolate the quadrupole. The
polarization induced by this strain is \citep{2016ARA&A..54..227K}
\begin{align}
    \label{e:pn}
    \Pp_{ab}(\pru n)
    &=
    2
    \int_0^{\chi_i} d\chi
    \int \frac{d^3\vecK}{(2\pi)^3}
    S^{(T)}_P (K, \chi)
    \hp_{ab}(\pru n,\chi, \vecK)
    .
\end{align}
Here, $S^{(T)}_P (K, \chi)$ is the
polarization source function for tensors, whose definition is given in
\citet{1996ApJ...469..437S}. It includes all the Boltzmann physics of how
gravity waves induce CMB polarization. Note that it includes the transfer
function of the gravity waves and $h_{ab}$ is understood to represent the
primordial value. Crucially, $S^{(T)}_P$ depends only on the magnitude of the
wave vector, $K$, and on our time coordinate $\chi$, \textit{i.e.} on the
evolution of the tensor mode and photons. The geometrical
dependence of the polarization (dependence on $\pru n$ and $\pru K$) is encoded
in $\hp_{ab}$.

On a flat patch of sky centred on the $\pru n = \pru z$ direction,
Equation~\ref{e:pn} can be Fourier transformed to be a function
of $\prv \ell$, the variable Fourier conjugate to $\pru n$. The Fourier
transform picks out only modes with $\delta^\perp_{ab} K^b = \ell_a / \chi$ as
contributing to each $\prv \ell$, and we define $\vecK_\ell \equiv \vecl /\bar
\chi + \pru z K_\parallel$. In this space the $E$-
and $B$-mode polarization tensors have simple forms:
\begin{align}
        e^{\perp E}_{ab}(\pru \ell)
    &=
    \sqrt{2}
    \left(
        \hat \ell_a \hat \ell_b
        - \frac{1}{2}\delta^{\perp}_{ab}
    \right)
    \nonumber\\
    e^{\perp B}_{ab}(\pru \ell)
    &=
    \frac{\sqrt{2}}{2}
    \left(
        \hat \ell_a \hat e^{\phi_\ell}_b
        + \hat e^{\phi_\ell}_a\hat \ell_b
    \right)
    ,
\end{align}

Decomposing $\Pp_{ab}(\vecl)$ into $E$ and $B$ modes as
$
\Pp_{ab}
= 
e^{\perp E}_{ab} a_E
+ e^{\perp B}_{ab} a_B
$,
and noting that the projections of the gravity-wave polarization tensors onto
the $E$- and $B$-mode polarization tensors can be written solely in terms of
$\mu_{K_\ell}\equiv K_\parallel / K_\ell$, we find
\begin{align}
    \label{e:Cl_EB}
    C^{EB}(\ell)
    &=
    \int_0^{\chi_i}
    \frac{
        d\chi
        d\chi'
    }{
        \bar \chi^2
    }
    \int_{-\infty}^{\infty} \frac{dK_\parallel}{2\pi}
    S^{(T)}_P (K, \chi)
    S^{(T)}_P (K, \chi')
    \nonumber\\
    &~~~\times
    \sin{[(\chi - \chi') K_\parallel]}
    (1 +\mu_{K_\ell}^2)
    \mu_{K_\ell}
    \heliP(K_\ell)
    P_h(K_\ell)
    .
\end{align}
It is seen that for thin last scatter---that is if the source function
$S^{(T)}_P$ approximates a delta function at recombination, $\chi = \chi_*$,
which was assumed in our earlier argument---then
the sine factor is zero and the correlations vanish. However, even if
this is not the case (\textit{i.e} if the effects of reionization are included),
the integrand is antisymmetric under exchange of $\chi$
and $\chi'$, while the integration limits are symmetric,
and the correlations still vanish.

To compute $C^{TB}(\ell)$, one replaces the factor of
$1 + \mu_{K_\ell}^2$ with $-1 + \mu_{K_\ell}^2$
in Equation~\ref{e:Cl_EB}. One also replaces
one factor of $S^{(T)}_P$ with
$S^{(T)}_T$, which has an extra contribution proportional to the time
derivative of the tensor transfer function
$\transfer(K,\chi)$. This is because gravity wave strain induces temperature
anisotropies both through Thompson scattering and through direct redshifting of
CMB photons. This breaks the $\chi$, $\chi'$ antisymmetry yielding
nonvanishing correlations between $B$-modes induced by Thompson scattering and
temperature modes induced by direct redshifting.

\blue{We note that this derivation applies equally well for vector modes,
replacing $S^{(T)}_P$ and $S^{(T)}_T$ with the appropriate source functions:
$S^{(V)}_P$ and $S^{(V)}_T$. In
this case we expect the $TB$ correlations to be highly suppressed by the sine
factor for mechanisms where the generation of anisotropies is confined to the
last-scattering surface.}

That the $EB$ correlations vanish in flat sky would seem to conflict with the
full-sky calculation that finds not only nonvanishing $EB$ correlations (at the
$0.01 \times \heliP$ level) but no
$1/\ell$ suppression of $EB$ compared to $BB$ and $EE$ from tensors. We find that
this is a direct result of sky curvature. In the Supplementary Material we
convert the full-sky $EB$ power spectrum directly to the flat-sky expression
above.
This conversion includes the regular flat-sky transformation
\begin{align}
    &\frac{2}{\pi}\int_0^\infty K^2 dK F(K, \chi, \chi') j_\ell(K\chi) j_\ell(K\chi')
    \nonumber\\
    &\qquad\qquad\rightarrow
    \int_{-\infty}^\infty \frac{dK_\parallel}{2\pi} F(K_\ell, \chi, \chi')
    e^{iK_\parallel(\chi - \chi')},\nonumber
\end{align}
where $F$ may include differential operators with respect to $\chi$ and
$\chi'$. This conversion makes approximations that are invalid before the first
few oscillations of the Bessel functions, at $K\chi \sim \ell + 1/2$---the
contribution to the integral from the nonplanar nature of the sky. While
this part of the integral is normally sub-dominant, the induced error is
unsuppressed by $1/\ell$. For the $EB$ correlations, the rest of
the integral vanishes, leaving sky sphericity to dominate the total signal.
That the first few oscillations of the Bessel functions dominate
is consistent with the findings of \citet{2007JCAP...09..002S} who
directly plotted the integration kernels.

\prlsection{Large-scale structure}

In the work of \citet{2010PhRvL.105p1302M} an effect was identified whereby large-scale
tensor perturbations tidally imprint local anisotropy in the smaller-scale
distribution
of matter. This imprint persists indefinitely, even after the tensor mode itself
has decayed by redshifting, and thus constitutes a fossilized map of the
primordial
tensor field. A more complete treatment of the fossil effect was performed by
\citet{2013PhRvD..88d3507D} and \citet{2014PhRvD..89h3507S} who identified
extra contributions and fully treated the dynamics. That effects of this kind
could be used to search for parity violation was first suggested by
\citet{2012PhRvL.108y1301J}, who dealt with more general second order
couplings between matter and extra fields rather than specifically 
the tidal interaction from gravity waves. Here we calculate the sensitivity of
large-scale structure surveys to gravity-wave helicity through tidal fossils.

The effect of large-scale tensor perturbations on the statistics of the
smaller-scale matter field is given by
\begin{align}
    &\langle \delta(\veck)  \delta({\veck}')\rangle \big|_{h_{ab}} =
    P(\veck) \bigg\{
        (2\pi)^3\delta^3(\veck+{\veck}')
        \nonumber\\
        &\quad+
        h^{ab}(\veck+{\veck}')\hat k_a \hat k_b'
        \left[\frac{1}{2}\frac{d \ln P}{d\ln k} -
        2S(|\veck+{\veck}'|)\right]
    \bigg\}.
        \label{e:fossils}
\end{align}
Here, $\langle\rangle|_{h_{ab}}$ represents an ensemble average over
realizations of the matter field while holding the tensor perturbations
$h_{ab}$ constant.
The equation is valid in the squeezed limit---where
$K = |\veck+\veck'| \ll k$. Throughout it is to be
understood that the tensor field, $h^{ab}$, is the primordial value,
whereas the matter field, $\delta(\veck^s)$, is evaluated at the epoch being observed.
The function $S(K)$ describes the growth of the tidal interaction and is
cosmology dependent \citep{2014PhRvD..89h3507S}.
It is of order unity and for the most relevant scales (modes
entering the horizon during matter domination) it
is roughly equal to $0.4$. Tensor modes are tidally imprinted on the
large-scale structure as the
gravity waves decay, and as such the above expression is valid for $K > aH$, as
larger scales have not yet begun to evolve. \blue{In addition, nonlinear evolution
will isotropise the density perturbations on small scales, making this
expression valid only for $k$ in the linear regime.}

To extract the information from this effect, a quadratic estimator on the matter
field is
used to form a noisy map of the tensor field, $\widehat{h_{ab}(\veck)}$.
\citet{2012PhRvL.108y1301J} described this procedure in detail, which we adapt
in the Supplementary Material and outline here. The optimal estimator is
\begin{align}
    \label{e:estimator}
    \widehat{h_{ab}(\vecK)} =
    \frac{N_h(K)}{4}
    &\sum_{\veck}
    \frac{
        w_{abcd}(\pru{K})
        \hat k^c \hat k^d f(k, K)
    }{
    2VP^{\rm tot}(k)P^{\rm tot}(|\vecK - \veck|)
    }
    \nonumber\\
    &\times\delta(\veck)  \delta(\vecK - \veck)
    ,
\end{align}
with
\begin{equation}
    f(k, K) \equiv P(k)\left[-\frac{1}{2}\frac{d \ln P}{d\ln k} +
    2S(K)\right].
\end{equation}
The tensor noise power spectrum is
\begin{equation}
    \label{e:noise}
    N_h(K) =
    \left\{
    \frac{1}{8}
    \sum_{\veck}
    \frac{
        [1 - (\hat K^a \hat k_a)^2]^2
        f(k, K)^2
    }{
        2VP^{\rm tot}(k)P^{\rm tot}(|\vecK - \veck|)
    }
    \right\}^{-1}
    .
\end{equation}
In the above equations, $P^{\rm tot}(k)$ is the power spectrum of the matter field
including any
noise, and we have ignored the difference between $\pru k$ and $\widehat{\veck -
\vecK}$ since we are working in the squeezed limit.
It is seen that in
three dimensions, estimates can be made for the tensor field on a mode-by-mode
basis that include all the geometrical tensor structure of $h_{ab}$. Thus, we
expect estimates of helicity to be limited only by the noise
on the gravity waves themselves, not by contamination from fields with
different tensor structure.
\blue{This also prevents contamination of the
tensor signal with other sources of shear, such as density--density tides
\citep{2012arXiv1202.5804P, 2016PhRvD..93j3504Z}
and weak gravitational lensing \citep{2004NewA....9..417P, 2004NewA....9..173C}.}

Estimators for the tensor power spectrum $\widehat{P_h(\vecK)}$ and helicity
power spectrum $\widehat{\heliP(\vecK)P_h(\vecK)}$ can be formed using the
contractions of $\widehat{h_{ab}(\veck)}$ from Equation~\ref{e:contractions}.
The uncertainties on these power spectrum estimators can be obtained by Wick
expanding the four-point functions, yielding
\begin{align}
    &\Cov(
        \widehat{P_h(\vecK)},
        \widehat{P_h(\vecK')}
        )
    =
    \Cov(
        \widehat{\heliP(\vecK)P_h(\vecK)},
        \widehat{\heliP(\vecK')P_h(\vecK')}
        )
    \nonumber\\
    &\qquad\qquad=
    \frac{1}{2}
    \left[
        \delta_{\vecK,-\vecK'}
        +
        \delta_{\vecK,\vecK'}
    \right]
    \left[P_h(K) + N_h(K)\right]^2,
    \nonumber\\
    &\Cov(
        \widehat{P_h(\vecK)},
        \widehat{\heliP(\vecK') P_h(\vecK')}
        )
    = 0.
\end{align}
If $\heliP$ is scale independent, the above equations
show that the uncertainties on $P_h$ and $\heliP P_h$ are equal and uncorrelated.
If $|\heliP|$ takes its maximal value of unity then detecting helicity has the
same difficulty as detecting the tensor modes in the first place. If $|\heliP|$
is small then its uncertainty is the reciprocal signal-to-noise ratio of
the tensor power.

We adopt the standard inflationary form for the tensor power spectrum
$P_h(K) = 2 \pi^2 r A_s/K^3$, where the tensor-to-scalar ratio $r$ is the only
free parameter.
We further assume that the factor
$(-1/2)({d \ln P}/{d\ln k}) + 2S$ is constant for large $k$ (which dominates
the information) and for $K>aH(z)$
(where Equation~\ref{e:fossils} is valid) and zero otherwise. The tensor noise power
spectrum is then
\begin{equation}
    N_h \approx  \frac{45 (2\pi)^2}{k_{\rm max}^3}
    \left(-\frac{1}{2}\frac{d \ln P}{d\ln k} + 2S\right)^{-2},
\end{equation}
and the uncertainties on final parameters are
\begin{align}
    \sigma_{(\heliP r)}
    &=
    \sigma_r
    \nonumber\\
    &=
    \left[
        \sum_{\vecK}
        \left(\frac{P_h(K)}{r}\right)^2
        \left(\frac{1}{P_h(K) + N_h}\right)^2
    \right]^{-1/2}
    \nonumber\\
    &\approx
    \frac{N_h}{2\pi A_s}
    \left(
        \frac{6 K_{\rm min}^3}{V}
    \right)^{1/2}
    \left[1 + \frac{P_h(K_{\rm min})}{N_h}\right]^{1/2}.
\end{align}
The last factor in the above expression is a correction for the sample variance
of the tensor field, which turns out to be significant even for $r \sim
\sigma_r$ due to the redness of $P_h$.

\blue{We now determine what
survey parameters $V$ and $k_{\rm max}$ are required to detect helicity at
$3\sigma$ significance for various values of $r$ and $|\heliP|$.}
Setting ${d \ln P}/{d\ln k} \approx -2.75$, $S\approx0.4$, $A_s=2.12\times
10^{-9}$ \citep{2016A&A...596A.107P}, and assuming a $z>10$ survey
of dark-ages structure when $aH\approx 1\,h/{\rm Gpc}^{-1}$, 
we find that if $r$ is at its
current upper limit of $r=0.07$ \citep{2016PhRvL.116c1302B} then a
$V=200\,({\rm Gpc}/h)^3$ survey measuring scales down to
$k_{\rm max}=7.6\,h/{\rm Mpc}$
could detect maximal helicity.
If instead $r=10^{-4}$, a survey of the same volume would need to measure
scales down to $k_{\rm max}=67\,h/{\rm Mpc}$,
although the same survey could detect $|\heliP| = 1.1\%$ if $r$ is
at the current upper limit.
As noted by \citet{2010PhRvL.105p1302M} and \citet{2012PhRvL.108y1301J},
such measurements are futuristic but
within the limits of what might be achievable through 21\,cm
surveys of prereionization structure.
\blue{The cosmic variance limit of observable helicity
is primarily set by the smallest scale that contains
information. At very high redshift, this is the Jean's scale, below which the
hydrogen gas does not cluster due to pressure
support \citep{2004PhRvL..92u1301L}.
In the $30<z<120$ range, which contains roughly $1000\,({\rm Gpc}/h)^3$ of
comoving volume, this scale is $k_{\rm max}\sim200\,h/{\rm Mpc}$. A survey
capturing all this information could detect maximal helicity if $r\sim10^{-7}$,
or $|\heliP| = 10^{-4}$ if $r$ is
at the current upper limit. We note that such a survey would require a
telescope several thousand kilometres in extent which would likely have to be
located in space.
}

Detecting helicity in primordial gravity waves would be a direct indication of
parity-violating physics in the very early Universe. Unfortunately, 
two-dimensional probes such as the CMB anisotropies are largely insensitive to the
helicity. The projection to two dimensions has two effects: suppressing the
signal due to approximate reflective symmetries, and confusing the tensor-like
modes with scalar modes, leading to additional noise contributions. In
contrast, three-dimensional probes allow the handedness of gravity waves
to be determined on a mode-by-mode basis, alleviating both of these issues. As
such, mapping gravity waves using their fossilized tidal imprints in the
large-scale structure could permit percent level helicity to be detected.

\blue{
In the modern Universe, binary systems emit gravity waves with opposite circular
polarization
above and below orbital plane.
If gravity were parity violating one would expect an asymmetric emission
of radiation, resulting in the net transfer of
momentum to the binary.
Such scenarios are constrained by the Laser
Interferometer Gravitational-Wave Observatory events
\citep{2016PhRvL.116x1102A, 2016PhRvL.116v1101A} as well as pulsar timing
\citep{2006Sci...314...97K} which show no deviations from general relativity.
Binary systems, however, probe a completely different physical regime than the
early Universe, and so constitute a complementary probe of gravitational
parity.
}

Looking forward, direct detection experiments could probe primordial gravity
waves on scales ranging from centimetres to
light-years.
For direct detection, time
dependence provides additional dimensionality and thus geometrical information.
Already, bounds from pulsar
timing arrays are allowing us to constrain some scenarios of black hole
formation~\cite{Pen:2015qta,Nakama:2016gzw}; \blue{however timing arrays
are unable to discern helicity for an
isotropic background \cite{2016PhRvD..93f2003K}}.

The present work emphasizes that
the detailed statistical properties of a stochastic gravity-wave
background may
in time become a vital source of new information about fundamental physics.

\begin{acknowledgments}
We thank Donghui Jeong, Marc Kamionkowski, Gary Hinshaw, and 
Mark Halpern for valuable discussions.
K.W.M.~is supported by the Canadian Institute for Theoretical Astrophysics
National Fellows program.
U.-L.P.~acknowledges support from the Natural Sciences and Engineering Research Council of Canada.
Research at Perimeter Institute is supported by the Government of Canada through the Department of Innovation, Science and Economic Development Canada  and by the Province of Ontario through the Ministry of Research, Innovation and Science. 
\end{acknowledgments}

\bibliography{gw_helicity}

\onecolumngrid
\newpage
\section{Supplementary Material}
\subsection{Parity violating CMB correlations in the flat-sky approximation}

Here we present the full derivation of the $EB$ correlations within the
flat-sky approximation. We begin by combining
Equations~\ref{e:projected_strain}~and~\ref{e:pn} and applying the flat
sky approximation. In flat sky, $\pru n = \pru z + \prv \theta$, and
$\delta^{\perp}_{ab} = \delta_{ab} - \hat z_a \hat z_b$. Additionally, the flat
sky approximation decouples the radial distance to structures, $\chi$, from the
distance use to convert angles to transverse distances, $\bar \chi$. This yields
\begin{align}
    \Pp_{ab}(\prv \theta)
    &=
    2
    \left[
    \delta^{\perp}_{ac}\delta^{\perp}_{bd}
    -\frac{1}{2} \delta^{\perp}_{ab}\delta^{\perp}_{cd}
    \right]
    \int_0^{\chi_i} d\chi
    \int \frac{d^3\vecK}{(2\pi)^3}
    S^{(T)}_P(k,\chi)
    e^{i(\bar \chi \prv \theta \cdot\vecK_\perp + \chi K_\parallel)}
    h^{cd}(\vecK)
    ,
\end{align}
which in harmonic space becomes
\begin{align}
    \Pp_{ab}(\vecl)
    &=
    \int d^2\prv \theta
    e^{-i\vecl\cdot\prv \theta}
    \Pp_{ab}(\prv \theta)
    \\
    &=
    2
    \left[
    \delta^{\perp}_{ac}\delta^{\perp}_{bd}
    -\frac{1}{2} \delta^{\perp}_{ab}\delta^{\perp}_{cd}
    \right]
    \int_0^{\chi_i} d\chi
    \int_{-\infty}^{\infty} \frac{dK_\parallel}{2\pi}
    S^{(T)}_P(k,\chi)
    e^{i\chi K_\parallel}
    h^{cd}(\vecK_\ell)
    .
\end{align}
Here $\vecK_\ell \equiv \vecl /\bar \chi + \pru z K_\parallel$.
In harmonic space we can define E- and B-mode polarization tensors. These are
\begin{align}
    e^{\perp E}_{ab}(\pru \ell)
    &=
    \sqrt{2}
    \left(
        \hat \ell_a \hat \ell_b
        - \frac{1}{2}\delta^{\perp}_{ab}
    \right)
    \\
    e^{\perp B}_{ab}(\pru \ell)
    &=
    \frac{\sqrt{2}}{2}
    \left(
        \hat \ell_a \hat e^{\phi_\ell}_b
        + \hat e^{\phi_\ell}_a\hat \ell_b
    \right)
    ,
\end{align}
who obey orthogonality relation
\begin{align}
    e^{\perp \beta}_{ab}(\pru \ell) e^{\perp \gamma\,ab}(\pru \ell) =
    \delta^{\beta\gamma}.
\end{align}

Decomposing $\Pp_{ab}(\vecl)$ into E- and B-modes as
$
\Pp_{ab}
= 
e^{\perp E}_{ab} a_E
+ e^{\perp B}_{ab} a_B
$,
and noting that
$
e^{\perp E}_{ab}(\pru \ell) e^{R\,ab}(\pru K_\ell)
=
e^{\perp E}_{ab}(\pru \ell) e^{L\,ab}(\pru K_\ell)
=
\frac{1}{2}(1 + \mu_{K_\ell}^2)
$
and
$
e^{\perp B}_{ab}(\pru \ell) e^{R\,ab}(\pru K_\ell)
=
- e^{\perp B}_{ab}(\pru \ell) e^{L\,ab}(\pru K_\ell)
=
i\mu_{K_\ell}
$,
we find
\begin{align}
    a_E(\vecl)
    &=
    \int_0^{\chi_i} d\chi
    \int_{-\infty}^{\infty} \frac{dK_\parallel}{2\pi}
    S^{(T)}_P(k,\chi)
    e^{i\chi K_\parallel}
    (1 + \mu_{K_\ell}^2)
    \left[h_R + h_L\right](\vecK_\ell)
    \\
    a_B(\vecl)
    &=
    \int_0^{\chi_i} d\chi
    \int_{-\infty}^{\infty} \frac{dK_\parallel}{2\pi}
    S^{(T)}_P(k,\chi)
    e^{i\chi K_\parallel}
    i
    2
    \mu_{K_\ell}
    \left[h_R - h_L\right](\vecK_\ell)
\end{align}
The $EB$ angular power spectrum is then
\begin{align}
    \langle a_E(\vecl)a_B(\vecl') \rangle 
    &=
    (2\pi)^2 \delta(\vecl + \vecl') C_{EB}(\ell)
    \\
    &=
    i2
    \int_0^{\chi_i} d\chi
    d\chi'
    \int_{-\infty}^{\infty} \frac{dK_\parallel}{2\pi}
    \frac{dK_\parallel'}{2\pi}
    S^{(T)}_P(k,\chi)
    S^{(T)}_P(k,\chi')
    e^{i\chi K_\parallel}
    e^{i\chi' K_\parallel'}
    (1 + \mu_{K_\ell}^2)
    \mu_{K_\ell}'
    \nonumber\\&~~~\times
    \left\langle
    \left[h_R + h_L\right](\vecK_\ell)
    \left[h_R - h_L\right](\vecK_\ell)
    \right\rangle
    \\
    &=
    i
    \int_0^{\chi_i} d\chi
    d\chi'
    \int_{-\infty}^{\infty} \frac{dK_\parallel}{2\pi}
    \frac{dK_\parallel'}{2\pi}
    S^{(T)}_P(k,\chi)
    S^{(T)}_P(k,\chi')
    e^{i\chi K_\parallel}
    e^{i\chi' K_\parallel'}
    (1 + \mu_{K_\ell}^2)
    \mu_{K_\ell}'
    \nonumber\\&~~~\times
    (2\pi)^3 \delta^3(\vecK_\ell + \vecK_\ell')
    \heliP(K_\ell)
    P_h(K_\ell)
    \\
    &=
    -i
    \int_0^{\chi_i} d\chi
    d\chi'
    \int_{-\infty}^{\infty} \frac{dK_\parallel}{2\pi}
    S^{(T)}_P(k,\chi)
    S^{(T)}_P(k,\chi')
    e^{i(\chi - \chi') K_\parallel}
    (1 + \mu_{K_\ell}^2)
    \mu_{K_\ell}
    \nonumber\\&~~~\times
    \frac{(2\pi)^2}{\bar \chi^2} \delta^2(\vecl + \vecl')
    \heliP(K_\ell)
    P_h(K_\ell)
\end{align}
We identify $\bar \chi$ as $(\chi + \chi')/2$.
As expected,
this parity violating correlation is sourced by the helicity
spectrum $\heliP P_h$. With the exception of the exponential factor, the integrand
is odd with respect
to $K_\parallel$ (or equivalently $\mu_{K_\ell}$).
This pulls out the imaginary part of the exponential factor:
\begin{align}
    C^{EB}(\ell)
    &=
    \int_0^{\chi_i}
    \frac{
        d\chi
        d\chi'
    }{
        \bar \chi^2
    }
    \int_{-\infty}^{\infty} \frac{dK_\parallel}{2\pi}
    S^{(T)}_P(k,\chi)
    S^{(T)}_P(k,\chi')
    \sin{[(\chi - \chi') K_\parallel]}
    (1 + \mu_{K_\ell}^2)
    \mu_{K_\ell}
    \heliP(K_\ell)
    P_h(K_\ell)
    .
    \nonumber
\end{align}
This is Equation~\ref{e:Cl_EB}.

\subsection{Direct conversion of curved-sky CMB to flat-sky}

The curved sky expectation of $C^{EB}(\ell)$ is \citep{2007JCAP...09..002S}
\begin{align}
    C^{EB}(\ell)
    &\approx
    -
    \frac{2}{\pi}
    \int_0^{\infty} dK K^2
    \frac{\heliP(K)P_h(K)}{2}
    \int_0^{\chi_i} d\chi
    S^{(T)}_P(K,\chi)
    \left[
    - j_\ell(K\chi)
    + j''_\ell(K\chi)
    \right]
    \int_0^{\chi_i} d\chi'
    S^{(T)}_P(K,\chi')
    2j'_\ell(K\chi'),
\end{align}
noting that the definition of $P_h(K)$ in
\textit{e.g.} \citet{1997PhRvD..55.1830Z} differs from the one used here by a
factor of ${(2\pi)^3}/{2}$. We have dropped terms that are suppressed by $\sim 1/\ell$.

Inverting the order of integration and rewriting the derivatives,
\begin{align}
    C^{EB}(\ell)
    &\approx
    -
    \frac{2}{\pi}
    \int_0^{\chi_i} d\chi d\chi'
    \int_0^{\infty} dK K^2
    \frac{\heliP(K)P_h(K)}{2}
    S^{(T)}_P(K,\chi)
    S^{(T)}_P(K,\chi')
    \left(
    -1
    + \frac{1}{K^2} \frac{d^2}{d\chi^2}
    \right)
    \left(
    \frac{2}{K} \frac{d}{d\chi'}
    \right)
    j_\ell(K\chi)
    j_\ell(K\chi'),
\end{align}

Following \citet{2007PhRvD..76h3005L}, for
$r > \nu + \nu^{1/3}/2$, and $\nu\equiv \ell + 1/2$ we have
\begin{equation}
j_\ell(r) \approx
\frac{
\sin\left[\sqrt{r^2 -\nu^2} - \arccos(\nu/r)/\nu + \pi /4\right]
}{
r (1 - \nu^2/r^2)^{1/4}
}.
\end{equation}
This approximation is only valid after the
first $\nu^{1/3}/2 \sim {\rm few}$ oscillations of the
Bessel function, meaning the contributions of these
first few oscillations to the integral are not properly
represented in the flat-sky approximation.

In applying the above approximation, we identify the expression
$\sqrt{K^2\chi^2 - \nu^2}$ to be
$\chi\sqrt{K^2 - \nu^2/\bar\chi^2} = \chi |K_\parallel|$. As above, we have used
the flat sky approximation to decouple the radial distance to structures,
$\chi$, from the distance used to project transverse distances to angles,
$\bar\chi$. A consequence is that the derivatives with respect to
$\chi$ and $\chi'$ act only on $\chi - \chi'$, not on $\bar\chi$.

We use the sine product formula to combine the two
approximations to Bessel functions, neglecting the term that oscillates 
rapidly as a function of $K$.
The $\arccos$ phase terms nearly
cancel and in any case contribute negligible phase.
We end up with
\begin{align}
    j_\ell(K\chi)
    j_\ell(K\chi')
    &\approx
    \frac{
        \cos
        \left[
            K_\parallel
            (\chi - \chi')
        \right]
    }{
        2 \bar\chi^2 K K_\parallel
    }
\end{align}

Changing variables of integration from $K$ to $K_\parallel$, we have
\begin{align}
    C^{EB}(\ell)
    &\approx
    -
    \frac{2}{\pi}
    \int_0^{\chi_i} d\chi d\chi'
    \int_{\nu/\bar\chi}^{\infty} dK K^2
    \frac{\heliP(K)P_h(K)}{2}
    S^{(T)}_P(K,\chi)
    S^{(T)}_P(K,\chi')
    \left(
    -1
    + \frac{1}{K^2} \frac{d^2}{d\chi^2}
    \right)
    \left(
    \frac{2}{K} \frac{d}{d\chi'}
    \right)
    \frac{
        \cos
        \left[
            K_\parallel
            (\chi - \chi')
        \right]
    }{
        2 \bar\chi^2 K K_\parallel
    }
    \\
    &\approx
    -i
    \int_0^{\chi_i}
    \frac{d\chi d\chi'}{\bar\chi^2}
    \int_{-\infty}^{\infty}
    \frac{
        dK_\parallel
    }{
        2\pi
    }
    \heliP(K_\ell)P_h(K_\ell)
    S^{(T)}_P(K_\ell,\chi)
    S^{(T)}_P(K_\ell,\chi')
    \left(
    1
    +\mu_{K_\ell}^2
    \right)
    \mu_{K_\ell}
    e^{
        i
        K_\parallel
        (\chi - \chi')
    }
    ,
\end{align}
which is the same as the flat sky derivation.
To calculate the $TB$ correlations, we replace
$- j_\ell + j''_\ell$ with $\nu^2j_\ell/(K\chi)^2 \approx j_\ell + j''_\ell$
(ultimately yielding $-1 + \mu_{K_\ell}^2$) and one factor of
$S^{(T)}_P$ with $S^{(T)}_T$.

\subsection{Fossil Estimators}

\citet{2012PhRvL.108y1301J} showed that the optimal estimator for the individual
polarization modes is
\begin{align}
    \widehat{h_\lambda(\vecK)} 
    &=
    N_\lambda(K)
    \sum_{\veck}
    \frac{
        e^{\lambda*}_{ab}(\pru{K})\hat k^a \hat k^b f^*(k, K)
    }{
    2VP^{\rm tot}(k)P^{\rm tot}(|\vecK - \veck|)
    }
    \delta(\veck)  \delta(\vecK - \veck),
    \\
    N_\lambda(K)
    &=
    \left[
    \sum_{\veck}
    \frac{|e^{\lambda*}_{ab}(\pru{K})\hat k^a \hat k^b f^*(k, K)|^2}
    {
    2VP^{\rm tot}(k)P^{\rm tot}(|\vecK - \veck|)
    }
    \right]^{-1}.
\end{align}
Here, $\lambda$ is one of $R$ or $L$ and
\begin{equation}
    f(k, K) \equiv  P(k)\left[-\frac{1}{2}\frac{d \ln P}{d\ln k} +
    2S(K)\right]
    .
\end{equation}
Note that our $f$ differs from the definition in \citet{2012PhRvL.108y1301J} by a
factor of $1/k^2$

We can re-expand this to an estimator for the tensor field:
\begin{align}
    \widehat{h_{ab}(\vecK)} &=
    \sum_\lambda \widehat{h_\lambda(\vecK)} e^{\lambda}_{ab}
    \\
    &=
    \sum_\lambda e^{\lambda}_{ab}(\pru{K}) N_\lambda(K)
    \sum_{\veck}
    \frac{
        e^{\lambda*}_{cd}(\pru{K})\hat k^c \hat k^d f^*(k, K)
    }{
    2VP^{\rm tot}(k)P^{\rm tot}(|\vecK - \veck|)
    }
    \delta(\veck)  \delta(\vecK - \veck)
    \\
    &=
    w_{abcd}(\pru{K})
    N_\lambda(K)
    \sum_{\veck}
    \frac{
        \hat k^c \hat k^d f^*(k, K)
    }{
    2VP^{\rm tot}(k)P^{\rm tot}(|\vecK - \veck|)
    }
    \delta(\veck)  \delta(\vecK - \veck).
\end{align}
Here we have used parity symmetry of the \emph{scalar} field to set
$N_R(K) = N_L(K) = N_\lambda(K)$.

The noise power spectrum can be simplified to
\begin{align}
    \left[N_h(K)\right]^{-1} &=
    \left[4N_\lambda(K)\right]^{-1}
    \\
    &=
    \frac{1}{4}
    \sum_{\veck}
    \frac{
        e^{\lambda}_{ab}(\pru{K}) e^{\lambda*}_{cd}(\pru{K})
        \hat  k^a \hat k^b \hat k^c \hat k^d|f^*(k, K)|^2
    }{
        2VP^{\rm tot}(k)P^{\rm tot}(|\vecK - \veck|)
    }
    \\ &=
    \frac{1}{8}
    \sum_{\veck}
    \frac{
        \left[
            e^{R}_{ab}(\pru{K}) e^{R*}_{cd}(\pru{K})
            + e^{L}_{ab}(\pru{K}) e^{L*}_{cd}(\pru{K})
        \right]
        \hat  k^a \hat k^b \hat k^c \hat k^d|f^*(k, K)|^2
    }{
        2VP^{\rm tot}(k)P^{\rm tot}(|\vecK - \veck|)
    }
    \\ &=
    \frac{1}{8}
    \sum_{\veck}
    \frac{
        w_{abcd}(\pru{K})
        \hat  k^a \hat k^b \hat k^c \hat k^d|f^*(k, K)|^2
    }{
        2VP^{\rm tot}(k)P^{\rm tot}(|\vecK - \veck|)
    }
    \\ &=
    \frac{1}{8}
    \sum_{\veck}
    \frac{
        [1 - (\hat K^a \hat k_a)^2]^2
        |f^*(k, K)|^2
    }{
        2VP^{\rm tot}(k)P^{\rm tot}(|\vecK - \veck|)
    }
    .
\end{align}

This yields Equations~\ref{e:estimator}~and~\ref{e:noise}:
\begin{align}
    \widehat{h_{ab}(\vecK)} &=
    w_{abcd}(\pru{K})
    \frac{N_h(K)}{4}
    \sum_{\veck}
    \frac{
        \hat k^c \hat k^d f^*(k, K)
    }{
    2VP^{\rm tot}(k)P^{\rm tot}(|\vecK - \veck|)
    }
    \delta(\veck)  \delta(\vecK - \veck)
    \nonumber\\
    N_h(K) &=
    \left\{
    \frac{1}{8}
    \sum_{\veck}
    \frac{
        [1 - (\hat K^a \hat k_a)^2]^2
        |f^*(k, K)|^2
    }{
        2VP^{\rm tot}(k)P^{\rm tot}(|\vecK - \veck|)
    }
    \right\}^{-1}
    .
    \nonumber
\end{align}

To obtain the bias and uncertainty on the power-spectrum estimators, we need the
two- and four-point functions of the tensor estimator,
which requires the Wick expansion
of the four-point function of the scalars:
\begin{align}
    \langle \widehat{h_{ab}(\vecK)}\widehat{h_{cd}(\vecK')}\rangle|_{h_{ab}=0}
    &=
    w_{ab}^{\p\p ef}(\pru{K})w_{cd}^{\p\p gh}(\pru{K'})
    \frac{N_h(K)N_h(K')}{16}
        \sum_{\veck\veck'}
        \frac{
            \hat k_e \hat k_f f^*(k, K)\hat k'_g \hat k'_h f^*(k', K')
        }{
            4V^2P^{\rm tot}(k)P^{\rm tot}(|\vecK - \veck|)
            P^{\rm tot}(k')P^{\rm tot}(|\vecK' - \veck'|)
    }
    \nonumber\\
    &\qquad\times
    \left[
        \langle \delta(\veck)  \delta(\vecK' - \veck') \rangle
        \langle \delta(\vecK - \veck) \delta(\veck') \rangle
        +
        \langle \delta(\veck) \delta(\veck') \rangle
        \langle \delta(\vecK - \veck) \delta(\vecK' - \veck') \rangle
    \right]
    \\
    &=
    w_{ab}^{\p\p ef}(\pru{K})w_{cd}^{\p\p gh}(\pru{K'})
    \frac{N_h(K)N_h(K')}{16}
        \sum_{\veck\veck'}
        \frac{
            \hat k_e \hat k_f f^*(k, K)\hat k'_g \hat k'_h f^*(k', K')
        }{
            4V^2P^{\rm tot}(k)P^{\rm tot}(|\vecK - \veck|)
            P^{\rm tot}(k')P^{\rm tot}(|\vecK' - \veck'|)
    }
    \nonumber\\
    &\qquad\times
    V^2\left[
        \delta_{\veck, -(\vecK' - \veck')}
        \delta_{\vecK - \veck, -\veck'}
        +
        \delta_{\veck, - \veck'}
        \delta_{\vecK - \veck, -(\vecK' - \veck')}
    \right]P^{\rm tot}(k)P^{\rm tot}(|\vecK - \veck|)
    \\
    &=
    V \delta_{\vecK, - \vecK'}
    w_{ab}^{\p\p ef}(\pru{K})w_{cd}^{\p\p gh}(\pru{K'})
    \frac{N_h(K)N_h(K')}{16}
        \sum_{\veck\veck'}
        \frac{
            \hat k_e \hat k_f f^*(k, K)\hat k'_g \hat k'_h f^*(k', K')
        }{
            4VP^{\rm tot}(k')P^{\rm tot}(|\vecK' - \veck'|)
    }
    \nonumber\\
    &\qquad\times
    \left(
        \delta_{\vecK - \veck, -\veck'}
        +
        \delta_{\veck, -\veck'}
    \right)
    \\
    &=
    V \delta_{\vecK, - \vecK'}
    w_{ab}^{\p\p ef}(\pru{K})w_{cd}^{\p\p gh}(\pru{K'})
    \frac{N_h(K)N_h(K)}{16}
        \sum_{\veck}
        \frac{
            \hat k_e \hat k_f \hat k_g \hat k_h f^*(k, K) f(k, K)
        }{
            2VP^{\rm tot}(k)P^{\rm tot}(|\vecK - \veck|)
    }
    \\
    &=
    V \delta_{\vecK, - \vecK'}
    w_{abcd}(\pru{K})
    \frac{N_h(K)N_h(K)}{32}
        \sum_{\veck}
        \frac{
            [1 - (\hat K^a \hat k_a)^2]^2 f^*(k, K) f(k, K)
        }{
            2VP^{\rm tot}(k)P^{\rm tot}(|\vecK - \veck|)
    }
    \label{e:subs_unit_sum}
    \\
    &=
    V \delta_{\vecK, - \vecK'} \frac{w_{abcd}(\pru{K})}{4}
    N_h(K)
    .
\end{align}
Here we have used the following identity:
\begin{align}
    \sum_{\prv k} g(k) w_{abef}(\vecK)w_{cdgh}(\vecK)
    \hat k^e \hat k^f \hat k^g \hat k^h
    =
    \frac{w_{abcd}}{2}\sum_{\veck} g(k) [1 - (\hat k^e \hat K_e)^2]^2
    ,
    \label{e:unit_sum}
\end{align}
for arbitrary function $g$. This can be shown in the continuum limit
where the sum is replaced by an integral.

We thus have
\begin{align}
    \langle \widehat{h_{ab}(\vecK)}\widehat{h_{cd}(\vecK')}\rangle
    &=
    \frac{V \delta_{\vecK, - \vecK'}}{4}
    \left[
        w_{abcd}(\pru{K})P_h(K)
        + w_{abcd}(\pru{K})N_h(K)
        + v_{abcd}(\pru{K})\alpha(K)P_h(K)
    \right]
    ,
\end{align}
which has contractions
\begin{align}
    \langle \widehat{h^{ab}(\vecK)}\widehat{h_{ab}(\vecK')}\rangle
    &=
    V \delta_{\vecK, - \vecK'}
    \left[P_h(K) + N_h(K)\right]
    \\
    \langle \widehat{h^{ab}(\vecK)}(-i\hat K^c\varepsilon_{ca}^{\p\p
    d})\widehat{h_{db}(\vecK')}\rangle
    &= 
    V \delta_{\vecK, - \vecK'} \alpha(K)P_h(K)
    .
\end{align}

From here it is straightforward to show that the four-point function of
the field estimator is
\begin{align}
    &
    \langle
        \widehat{h_{ab}(\vecK)}\widehat{h_{cd}(-\vecK)}
        \widehat{h_{ef}(\vecK')}\widehat{h_{gh}(-\vecK')}
    \rangle
    -
    \langle
        \widehat{h_{ab}(\vecK)}\widehat{h_{cd}(-\vecK)}
    \rangle
    \langle
        \widehat{h_{ef}(\vecK')}\widehat{h_{gh}(-\vecK')}
    \rangle
    \nonumber\\
    &\qquad
    =
    V^2 \frac{1}{16}
    \left[
        \delta_{\vecK,-\vecK'}w_{abef}(\vecK)w_{cdgh}(\vecK)
        +
        \delta_{\vecK,\vecK'}w_{abgh}(\vecK)w_{cdef}(\vecK)
    \right]
    \left[P_h(K) + N_h(K)\right]^2
    .
\end{align}
This has contractions
\begin{align}
    &
    \langle
        \widehat{h^{ab}(\vecK)}\widehat{h_{ab}(-\vecK)}
        \widehat{h^{cd}(\vecK')}\widehat{h_{cd}(-\vecK')}
    \rangle
    -
    \langle
        \widehat{h^{ab}(\vecK)}\widehat{h_{ab}(-\vecK)}
    \rangle
    \langle
        \widehat{h^{cd}(\vecK')}\widehat{h_{cd}(-\vecK')}
    \rangle
    \nonumber\\
    &\qquad
    =
    V^2 \frac{1}{2}
    \left[
        \delta_{\vecK,-\vecK'}
        +
        \delta_{\vecK,\vecK'}
    \right]
    \left[P_h(K) + N_h(K)\right]^2
    \\
    &
    (-i\hat K^c\varepsilon_{ca}^{\p\p d})
    (-i\hat K'^g\varepsilon_{ge}^{\p\p h})
    \left[
        \langle
            \widehat{h^{ab}(\vecK)}\widehat{h_{db}(-\vecK)}
            \widehat{h^{ef}(\vecK')}\widehat{h_{hf}(-\vecK')}
        \rangle
        \right.
        -
        \left.
        \langle
            \widehat{h^{ab}(\vecK)}\widehat{h_{db}(-\vecK)}
        \rangle
        \langle
            \widehat{h^{ef}(\vecK')}\widehat{h_{hf}(-\vecK')}
        \rangle
    \right]
    \nonumber\\
    &\qquad
    =
    V^2 \frac{1}{2}
    \left[
        \delta_{\vecK,-\vecK'}
        +
        \delta_{\vecK,\vecK'}
    \right]
    \left[P_h(K) + N_h(K)\right]^2
    \\
    &
    (-i\hat K'^g\varepsilon_{ge}^{\p\p h})
    \left[
        \langle
            \widehat{h^{ab}(\vecK)}\widehat{h_{ab}(-\vecK)}
            \widehat{h^{ef}(\vecK')}\widehat{h_{hf}(-\vecK')}
        \rangle
        \right.
        -
        \left.
        \langle
            \widehat{h^{ab}(\vecK)}\widehat{h_{ab}(-\vecK)}
        \rangle
        \langle
            \widehat{h^{ef}(\vecK')}\widehat{h_{hf}(-\vecK')}
        \rangle
    \right]
    = 0
\end{align}

\end{document}